\documentclass[12pt]{article}
\usepackage{graphicx}

\input{tcilatex}

\begin{document}

\title{Power Laws of Wealth, Market Order Volumes and Market Returns}
\author{Sorin Solomon \\
%EndAName
{\small {Racah Institute of Physics, Hebrew University of Jerusalem, Israel }%
}\\
Peter Richmond\\
{\small {Department of Physics, Trinity College Dublin, Ireland }}}
\date{}
\maketitle

\begin{abstract}
Using the Generalised Lotka Volterra (GLV) model adapted to deal with muti
agent systems we can investigate economic systems from a general viewpoint
and obtain generic features common to most economies. Assuming only weak
generic assumptions on capital dynamics, we are able to obtain very specific
predictions for the distribution of social wealth. First, we show that in a
'fair' market, the wealth distribution among individual investors fulfills a
power law. We then argue that 'fair play' for capital and minimal
socio-biological needs of the humans traps the economy within a power law
wealth distribution with a particular Pareto exponent $\alpha \sim 3/2$. In
particular we relate it to the average number of individuals L depending on
the average wealth: $\alpha \sim L/(L-1)$. Then we connect it to certain
power exponents characterising the stock markets. We obtain that the
distribution of volumes of the individual (buy and sell) orders follows a
power law with similar exponent $\beta \sim \alpha \sim 3/2$. Consequently,
in a market where trades take place by matching pairs of such sell and buy
orders, the corresponding exponent for the market returns is expected to be
of order $\gamma \sim 2 \alpha \sim 3$. These results are consistent with
recent experimental measurements of these power law exponents ([Maslov 2001]
for $\beta$ and [Gopikrishnan et al. 1999] for $\gamma$). \vskip 12pt \emph{%
Keywords}: Pareto-Zipf, Lotka-Volterra, power law; random multiplicative
process; wealth distribution; market returns.
\end{abstract}

%\begin{frontmatter}

\section{Power Laws and Logistic Equations}

\qquad

It was first observed by Pareto [1897 sic!] that in an economy, the fraction 
$P(w)$ of people owning a wealth $w$ is proportional to a power of $w$:

\begin{equation}
P(w)\sim w^{-1-\alpha }
\end{equation}
\qquad

For the last hundred years the value of $\alpha \sim $ 3/2 changed little in
time and across the various capitalist economies.

Aoki and Yoshikawa [1999] quote Montroll [1978] to the effect that ''almost
all the social phenomena, except in their relatively brief abnormal times
obey the logistic growth'' [Lotka 1925, Volterra 1926]: 
\begin{equation}
dw/dt=Aw-Bw^{2}
\end{equation}

It was shown [Solomon and Levy 1996, Levy and Solomon 1996, Biham et. al
1998] that Generalized Lotka Volterra (GLV) systems explicitating the
aggregated logistic equation (2) at the microscopic agents level lead to
power law distributions of the form (1). One can therefore say that the
careful reconsideration of the system (2), led to the solution of a 100 year
old puzzle by a 75 year old equation.

In the next sections we

- introduce the GLV model,

- show that it reduces to a set of decoupled stationary linear stochastic
equations.

- derive analytically the Pareto law for the relative wealth distribution in
the GLV model.

- explain the Pareto exponent $\alpha \sim $ 3/2 based on intrinsic
biological and social constraints

- relate via the GLV dynamics the Pareto exponent $\alpha \sim $ 3/2 to the
exponents of various stock market distributions. In particular GLV suggests
that distribution of the volumes of the individual buy/sell orders is a
power law with an exponent $\beta \sim 3/2$. For the distribution of the
trade-by-trade returns the predicted exponent is $\gamma \sim $ 3. \newline

\section{The Generalized Lotka -Volterra Model}

Let us consider the dynamical system:\qquad \qquad 
\begin{equation}
dw_{i}(t)=w_{i}(t+ \tau )-w_{i}(t)=[\varepsilon _{i}(t)\mathbf{\sigma }%
_{i}+c_{i}(w_{1},w_{2},...,w_{N},t)]w_{i}(t)+a_{i}\mathbf{\Sigma }%
_{j}b_{j}w_{j}(t)
\end{equation}
\noindent with $i,j=1,2,...,N$.

The interpretation of the various terms is as follows:

$\bullet $ $dw_{i}(t)=w_{i}(t+\tau )-w_{i}(t)$ is the time evolution of the
wealth of the individual $i$ during the time interval $\tau$.

$\bullet $ The random variables $\varepsilon _{i}(t)$ represent the purely
stochastic fluctuations in the returns of the various investors and since we
express the square standard deviations by the variables \textbf{$\sigma $}$%
_{i}^{2}$ one can assume without loss in generality that 
\begin{equation}
<\varepsilon _{i}(t)^{2}>=1
\end{equation}

$\bullet $ The systematic endogenous and exogenous trends in the returns are
expressed by the arbitrary functions $c_{i}(w_{1},w_{2},...,w_{N},t)$.

$\bullet $ Since the average value, $m_{i}(t)=<\varepsilon _{i}(t)>$ can be
absorbed in $c_{i}$, we may assume without loss in generality that: 
\begin{equation}
m_{i}(t)=<\varepsilon _{i}(t)>=0\ 
\end{equation}

$\bullet $ the terms $a_{i}\mathbf{\Sigma }_{j}b_{j}w_{j}(t)$ represent the
wealth redistribution by taxes, salaries, subsidies, etc. More precisely $%
b_{j}$represents how much of the wealth of the individual $j$ contributes to
the total wealth to be redistributed while the coefficients $a_{i} > 0$
represent the amount of wealth redistributed to the individual $i$.

Since an overall factor in the $b_{j}$'s can be absorbed in an overall
factor in the $a_{i}$'s, one can assume without loss in generality that:

\begin{equation}
\mathbf{\Sigma }_{j}b_{j}=1
\end{equation}

We will also assume here that all \textbf{$\sigma $}$_{i}^{2}$, $c_{i}$ and $%
a_{i} $ are of order $\tau $ and all $b_{j}$'s are positive of order $1/N$.
Situations in which the coefficients describe certain neighbourhood
structures in the investor's space have been studied in [Shnerb et. al 2000,
2001].

\bigskip

For arbitrary, unequal $c_{i}$ factors, the dynamics of (3) will lead to
systematically increasing inequalities in the \ relative wealths $x_{i}$. In
particular, individuals $i$ with very negative $c_{i}$will keep loosing
wealth. If they do it indefinitely, they will eventually disappear from the
market (see however Levy, Persky and Solomon 1996, Farmer 1999, Solomon and
Levy 2000, for other periodic, quasi-periodic and more complex regimes). The
disappearence of the weakest, together with a host of other phenomena
(adaptability, exploitation until exhaustion of systematic gain
opportunities, adverse market effects: the largest players bidding against
oneselves) leads usually to the ''\textbf{efficient market}'' regime. In
simple words, ''market efficency'' means that the market is more efficient
than any one individual in discovering the 'correct' prices and eliminating
thereby the possibility of systematic speculative gain. While it is
interesting to study the effects of eventual departures from ''market
efficiency'' and their duration [Farmer 1999, Levy, Perski and Solomon 1996,
Solomon and Levy 2000] it is reasonable to assume that for long time
intervals, market efficiency holds and that except for the purely stochastic
factors $\varepsilon _{i}(t)$ all the individuals have the same expected
relative returns. i.e: 
\begin{equation}
c_{i}(w_{1},w_{2},...,w_{N},t)=c(w_{1},w_{2},...,w_{N},t)
\end{equation}

Then, in the limit\newline

\begin{equation}
a_{i}/\mathbf{\sigma }_{i}^{2}>>1/lnN\newline
\end{equation}

one can show [Biham et al 1998, Huang and Solomon 2000, 2001, Blank and
Solomon 2000] that the global dynamics of the economy can be expressed by
taking the sum of the equations (3) each weighted by $b_{j}$ to obtain: 
\begin{equation}
du(t)=u(t+\tau )-u(t)=c(w_{1},w_{2},...,w_{N},t)u(t)+au(t)
\end{equation}
\noindent where 
\begin{equation}
a(t)=\mathbf{\Sigma }_{i}b_{i}a_{i}(t)
\end{equation}
\begin{equation}
u(t)=\mathbf{\Sigma }_{j}b_{j}w_{j}(t)
\end{equation}
\noindent and we have used equation (5) to neglect, in the limit $%
N\rightarrow \infty $, the sum of the random terms. One can think of the
variable $u$ as a weighted average wealth.

Obviously for arbitrary functions $c(w_{1},w_{2},...,w_{N},t)$, the time
evolution of $u(t)$ and the functions $w_{i}(t)$ can be very eventfull
[Levy, Persky and Solomon, 1996, Farmer 1999, Levy, Levy and Solomon 2000,
Solomon and Levy 2000]. However we show below that the probability
distribution of the relative wealth 
\begin{equation}
x_{i}(t)=w_{i}(t)/u(t)
\end{equation}
\noindent has sometimes a much simpler behavior. By applying the chain
differentiaton rule to (12) and using both (3) and (7) one gets: 
\begin{eqnarray}
dx_{i}(t) &=&dw_{i}/u-w_{i}/u^{2}du \\
&=&[\varepsilon _{i}(t)\mathbf{\sigma }_{i}+c]x_{i}(t)+a_{i}-x_{i}(t)[c+a]%
\newline
\\
&=&(\varepsilon _{i}(t)\mathbf{\sigma }_{i}-a)x_{i}(t)+a_{i}
\end{eqnarray}
I.e. the system of equations splits into a set of independent, linear, time
independent stochastic equations: 
\begin{equation}
dx_{i}(t)=\varepsilon _{i}\mathbf{\sigma }_{i}x_{i}(t)-ax_{i}(t)+a_{i}
\end{equation}
Note that this holds even if \textbf{$\sigma $}$_{i}$, $D_{i}$ and $a_{i}$
are functions of the corresponding $x_{i}$. In fact one obtains [Richmond
2000, Richmond and Solomon 2000, Solomon and Richmond 2000]: 
\begin{equation}
P(x_{i})=\frac{exp\{2\int [-ax_{i}(t)+a_{i}]/[\mathbf{\sigma }%
_{i}x_{i}]^{2}dx_{i}\}}{[\mathbf{\sigma }_{i}x_{i}]^{2}}
\end{equation}

In the particular case when \textbf{$\sigma $}$_{i}$ and $a_{i}$ are
constant, one obtains: 
\begin{equation}
P(x_{i})\sim x_{i}^{-1-\alpha _{i}}exp[\frac{-2a_{i}}{x_{i}\mathbf{\sigma }%
_{i}^{2}}]
\end{equation}
\noindent where 
\begin{equation}
\alpha _{i}=1+2a/\mathbf{\sigma }_{i}^{2}
\end{equation}

Note that in this approach one can have subpopulations that differ in their
regular additive incomes, $a_{i}$ and/or in their standard deviation for the
multiplicative (speculative) returns,$\mathbf{\sigma }_{i}^{2}$. In
particular if has, for example, a sub-population involved in a NASDAQ like
system with large $\mathbf{\sigma }_{i}^{2}$ this population will possess an 
$\alpha _{i}$ much smaller than the rest of the population engaged in
classical or 'old economy' investment behaviour. This in turn will imply the
concentration of wealth in a few 'lucky' hands as shown in section 4. This
will lead to market fluctuations which are significantly enhanced. This
simly reflects the fact that wealth changes that occur are fractions of
wealth of very wealthy individuals. (Think for example of the change in the
market index induced by Bill Gates losing the current court case.)
Obviously, a certain dependence of $a_{i}$ or \textbf{$\sigma $}$_{i}^{2}$
on $x_{i}$ may modify the power law exponent when substituted in (17).

In the case where all the coefficients are $i$-independent, i.e. $\mathbf{%
\sigma }_{i}^{2}=\mathbf{\sigma }^{2},a_{i}=a$ and $b_{i}=1/N$ one gets: 
\begin{equation}
P(x)\sim x^{-1-\alpha }exp[-2a/(\mathbf{\sigma }^{2}x)]
\end{equation}
\noindent with 
\begin{equation}
\alpha =1+2a/\mathbf{\sigma }^{2}
\end{equation}
[Kesten 1973, Solomon and Levy 1996, Sornette and Cont 1997, Solomon 1998,
Marsili, Maslov and Zhang 1998, Bouchaud and Mezard 2000]. The distribution $%
P(x)$ has a peak at $x_{0}=1/(1+D/a)$. Above $x_{0}$, the relative wealth
distribution $P(x)$ behaves like a power law while below $x_{0},$ $P(x)$
vanishes very fast.

One can show that for finite $N$, the main corrections are:

1) a factor which vanishes at $x=N$. This is consistent with the fact that
there cannot be an agent with wealth $w_{i}(t)$ larger than the total wealth 
$Nw(t)$: 
\begin{equation}
P(x)=x^{-1-\alpha }exp[-2a/(xD)]exp[-2a/(D(1-x/N))]
\end{equation}

2) a correction to $\alpha$: 
\begin{equation}
\alpha =1+2[a/D-K]/[1+K]
\end{equation}
\noindent where 
\begin{equation}
K=N^{-2+2/\alpha }\sim N^{-4a/D/(1+2a/D)}
\end{equation}

This implies $\alpha <1$ if $N<<exp(D/a)$ i.e. the wealth gets concentrated
in just a few hands [Malcai, Biham and Solomon 1999, Blank and Solomon 2000,
Huang and Solomon 2001].

\section{Why has the Pareto exponent been constant for the last 100 troubled
years}

Until now, we have explained the survival of the individual wealth power law
(18), (20), (22) in the presence of large time variations in the total
wealth. We now relate the constant value of $\alpha \sim 3/2$ measured over
the last 100 years (and for all the major capitalist economies) [Pareto
1897, Zipf 1949, Levy and Solomon 1997] to the social and biological
constraints imposed on any society. The main idea is to exploit the
particular characteristics of the shape of the wealth distribution curve
(20) in order to relate the power decay of the probability distribution at
large wealth to the wealth distribution of the poorest agents. This is
possible since both the exponent $\alpha =1+2a/\mathbf{\sigma }^{2}$ of the
power law for large wealth and the coefficient $2a/\mathbf{\sigma }^{2}$ in
the exponential of -1/x that dominates low wealth behavior, depend on the
single parameter $a/\mathbf{\sigma }^{2}.$ Consequently, how poor the poor
are allowed (or can afford) to be, determines the power low distribution of
more wealthy agents via the ratio $a/\mathbf{\sigma }^{2}$. The crucial
expression $a/\mathbf{\sigma }^{2}$ can be understood as the ratio between
additive regular incomes (salaries, social security, services) and the
purely random multiplicative incomes originating in speculative market
activities.

One of the crucial characteristics of the distribution $P(x)$ (20) is the
fact that as one goes to lower and lower $x \rightarrow 0$ values below it
maximum $x_0 \sim 1/(1+ D/a)$, its decay to 0 is extremely fast. In fact all
its derivatives are 0 at $x=0$. Given this sharp decay at low $w$ values,
one can assume that effectively, the distribution $P(x)$ vanishes below a
certain minimal relative wealth $x_{m}$ (poverty bound). The value of $x_{m}$
can be estimated roughly by assuming that there are no individuals below it
and that above it, the power law is fulfilled. Then one gets the relation
between $\alpha$ and x $_{m}$ from the identity 
\begin{eqnarray}
&<x>=<x_{i}(t)> \\
&=<w_{i}(t)/w(t)>=w/w=1
\end{eqnarray}
which implies: 
\begin{eqnarray}
1 & = & [\int_{x_{m}} x^{-\alpha }dx]/[\int_{x_{m}}x^{-1-\alpha }dx] \newline
\qquad \qquad \qquad \\
&=&[-1/(1-\alpha )x_{m}^{1-\alpha }]/[-1/(-\alpha )x_{m}^{-\alpha }]\newline
\qquad \qquad \qquad \\
&=&x_{m}\alpha /(\alpha -1)\newline
\end{eqnarray}
or: 
\begin{equation}
\alpha =1/(1-x_{m})
\end{equation}
\noindent [Ijiri and Simon 1977, Levy and Solomon 1996, Malcai et al 99,
Blank and Solomon 2000].

According (21) this yields:

\[
x_{m}=1-1/\alpha =1/(1+1/2 \ \mathbf{\sigma }^{2}/a) 
\]

Based on (30), (31) one can now give a general scenario of how the internal
interests and constraints within society lead to the actual value of $\alpha
\sim 3/2$ measured repeatedly in various economies in the last 100 years.
Suppose that in a given economy the wealth necessary to keep a person alive
is $K$. Certainly, anybody having less than that will have a very
destabilizing effect on the society, so the number of people with wealth
less then $K$ should be negligible if that society is to survive. Let us now
suppose that the average family supported by an average wealth, has in
average $L$ members . Clearly they will need a wealth of order $KL$,
otherwise the wage earners will seek to correct the situation by strikes,
negotiations, elections or revolts. Note that in a sense, the wealth of the
average family is the \textbf{definition} of the minimal amount for
supporting $L$ dependents, since the prices of the prime necessities will
always adjust to it (i.e. if the average wealth increases so will the prices
of housing, services, etc.).

In short, while the poorest people (who may not be able to afford a family)
will ensure for their own survival that they do not get less than $1/L$ of
the average, the average population will almost by definition take care that
their income is at least $L$ times the minimal wealth. In this way they
insure the survival of their offspring. All in all, we are lead to the
conclusion that 
\[
x_{m}\sim 1/L 
\]
is dictated by the compromise between the survival instincts of the
individual and the reproduction interesses of the species. Using equation
(30) one further predicts: 
\begin{equation}
\alpha =1/(1-x_{m})\sim L/(L-1) .
\end{equation}
These relations fit well the known numbers for typical capitalist economies
in the last century: family size $L\sim 3-4$, poverty line (below which
people get subsidized) 
\[
x_{m}\sim 1/4-1/3 
\]
and $\alpha \sim 1.33-1.5$.

Our other key conclusion is, therefore, that the lower bound on the relative
poverty governs totally the overall relative wealth distribution. The
details of the dynamics by which this distribution arises are of course
complex and depend on the interactions in the system.

Most likely, during periods of large speculative fluctuations ${\sigma }^{2}$
which can give rise to significant numbers of ''nouveau riches'' , regular
wage earners will demand that the salaries, pensions and social security
(all contributing to $a$ in) will increase in such a manner that the ratio $%
a/\mathbf{\sigma }^{2}$ returns to the value 1/6 - 1/4 consistent with $%
x_{m}=1/(1+1/2\mathbf{\sigma }^{2}/a)=1/L$ with $L \sim 3 - 4$. During
'bearish' periods (small ${\sigma }^{2}$) salaries $a$ will erode leading
again to a ratio $a/\mathbf{\sigma }^{2}$ that is fixed around: 
\[
a / {\sigma }^{2} \sim 1/[2(L-1)] \sim 1/4 - 1/6 
\]
. Consequently, the value of $\alpha \sim L/(L-1)$ is kept fixed too around $%
\alpha \sim 3/2 - 4/3 $. The Pareto experimental result $\alpha \sim 1.4$
falls well inside this range.

One may object that at the begining of the century the number of children
per couple was larger. However, one should keep in mind that at that time
children started to earn their living at an early age, so the number of
dependent children per wage earner was not far from 3-4. Moreover a slow
shift towards larger values of $\alpha$ (1.5-1.7) (corresponding to lower $L$
values) seems to take place in modern economies with strong social security
policies.

\section{\protect\bigskip Heavy Tails of Buy/ Sell Order Volumes and Market
Returns Distributions in GLV}

We discuss here some practical implications for the finacial markets of the
formal results obtained in the previous sections.

In particular we derive the values that GLV suggests for the exponents of
the probability distributions of

\begin{itemize}
\item  volumes of sell/buy orders = the amounts that the traders request to
buy or sell. GLV suggests an exponent of the order $\alpha \sim 3/2$.

\item  trade-by-trade returns = the stock price variations from one trade to
the next. GLV suggests and exponent $\gamma \sim 2\alpha \sim 3$.
\end{itemize}

According to Eq. (3), the variation of an individual investment is to lowest
order in $\tau $ the term proportional to \textbf{$\sigma $}$_{i}\sim \tau
^{1/2}$: 
\begin{equation}
dw_{i}(t)=w_{i}(t+\tau )-w_{i}(t)=\varepsilon _{i}(t)\mathbf{\sigma }%
_{i}w_{i}(t)
\end{equation}
That is, according the GLV model, the variations in the invested capital are
proportional to the capital itself. The proportionality of individual
investments and returns to the individual wealth is a crucial property of
the capital and a crucial ingredient of the GLV (and LLS) model. It is
consistent [Levy and Solomon 97] with the experimental fact that the
distribution of annual individual returns fulfills a power law with an
exponent $\alpha$ equal to the one of the wealth distribution ($3/2$).

In fact, if one uses the experimental fact that the investors have constant
relative risk aversion then, one can prove [Levy, Levy Solomon 2000] that
their utility functions are powers of their wealth. This in turn implies
that the fraction of wealth that such an investor considers as optimal to
invest is a fraction (that depends on the investor's current expectations)
of his personal wealth. This is to be contrasted with the implications of
the non-realistic (experimentally rejected but often used for analytical
expediency) constant absolute risk aversion assumption. This would lead to
an exponential utility function which predicts personal investments which
are independent on the investor's wealth.

It is therefore very reasonable to assume that each buy/ sell order volume
is stochastically proportional to the wealth of the trader that emits it.

Therefore, a natural implication of the GLV model [Levy and Solomon 97,
Solomon 2000] is that the volumes of the individual sell/buy orders are
distributed too by a power law with an exponent $\beta \sim \alpha \sim 3/2$
equal to the exponent of the Pareto individual wealth distribution. This is
confirmed by measurements reported by [Maslov 2001] and [Gopikrishnan et al.
2000] where the order volumes are distributed by a power law with exponent $%
\alpha < 2$.

Let us now estimate the exponent of the distribution of market price
variations from one trade to another (trade-by-trade returns). Since the
value of a stock is proportional to the sum of the individual investments in
that stock, the trade-by-trade returns are stochastically proportional to
the volumes traded in the individual trades. In fact the stochastic
correlations between trade volumes and absolute value of the returns are
well documented in the literature. In an ideal market where trading and
price fixing take place in a centralized way [LLS, Biham et al] via a market
maker who matches globally all individual demands / ofers such as to
saturate the overall market demand/ ofer, it follows that trade volumes (and
trade-by-trade returns) obey power laws with the same exponents as the buy/
sell order volumes $\gamma \sim \beta \sim \alpha \sim 3/2.$

However, in most real markets, the trades take place by the matching of
pairs of buy and sell orders with compatible prices. In such a case the
volume of each trade is equal to the smallest among the volumes of the
matched pair. The prediction for distribution of trade volumes is now very
different: Since the probability for each of 2 matched orders to exceed (or
equal) a certain volume $v$ is 
\begin{equation}
P(>v)\sim v^{-\alpha }
\end{equation}
\noindent the probability that both have a volume (equal or) larger than $v$
is the product: 
\begin{equation}
P(>v,>v)\sim P(>v)P(>v)\sim v^{-\alpha }v^{-\alpha }\sim v^{-2\alpha }
\end{equation}
So the prediction of GLV for such a market measurement is that the trades
volumes and the trade-by-trade returns will fulfill a power law with an
exponent 
\begin{equation}
\gamma \sim 2\alpha \sim 3
\end{equation}
While for very short times, and small returns the exponent $\gamma \sim 3$
in the returns power distribution is masked by various other effects (e.g.
subtelties related to the differentiation between market orders and limit
orders and the way one accounts for a large order that is saturated by the
sequential matching with a series of smaller orders; this accounts also for
some apparent discrepancies between [Maslov 2001] and [Gopikrishnan et. al
2000]), for a wide range of parameters, the empirical data confirm an
exponent $\gamma \sim 3$ in the power tails of the market returns
distributions [Gopikrishnan et al 1999].

\section{Conclusions}

By analyzing economic dynamics from a general viewpoint we have demonstrated
features that are common to most economies. Assuming only weak generic
assumptions on capital dynamics, we are able to obtain very specific
predictions for the distribution of social wealth. A crucial assumption is
that the capital market is fair, i.e. equal capitals have equal
opportunities. Mathematically, this is expressed via the assumption that
expectations for returns are independent of $i$ . We show that in such a
market, the wealth distribution among individual investors fulfills a power
law.The balance between ''fair play'' for the capital and minimal
socio-biological needs of the humans seems to trap the \ world economy into
a power law wealth distribution which determines much of its dynamical and
equilibrium properties and in particular the Pareto exponent $\alpha \sim
3/2 $. Moreover, the model is consistent with the recent measurments of the
power law exponents of the distributions of order volumes and market returns
[Maslov 2001, Gopikrishnan et al. 1999].

\bigskip

This paper is dedicated respectfully to the memory of Herbert Simon. \newline
\newline

\begin{center}
\textbf{References}
\end{center}

\noindent P. W. Anderson in The Economy as an Evolving Complex System II
(Redwood City, Calif.: Addison-Wesley, 1995), eds. W. B. Arthur, S. N.
Durlauf, and D. A. Lane. \newline
M. Aoki and H. Yoshikawa, Demand creation and economic growth, U. of Tokio ,
Ctr. for Int'l. Research on the Japanese Econ. 1999. \newline
O. Biham, O. Malcai, M. Levy, S. Solomon, Phys. Rev. E 58, 1352 (1998) 
\newline
A. Blank and S. Solomon "Power laws in cities population, financial markets
and internet sites (scaling in systems with a variable number of
components)" \newline
Physica A 287 (1-2) (2000) pp. 279-288. \newline
J. P. Bouchaud and M. M$\backslash $'ezard, Physica A 282, 536 (2000) 
\newline
Farmer J.D., "Market Force, Ecology and Evolution," e-print
adap-org/9812005. See also J.D. Farmer and S. Joshi, "Market Evolution
Toward Marginal Efficiency" SFI report 1999. \newline
P. Gopikrishnan, V. Plerou, X. Gabaix, H. E. Stanley Statistical Properties
of Share Volume Traded in Financial Markets; Phys. Rev. E. (Rapid Comm.), 62
(2000) R4493. \newline
P. Gopikrishnan, V. Plerou., L.A.N. Amaral, M. Meyer and H.E. Stanley, Phys.
Rev. E 60, 5305 (1999). \newline
Z. F. Huang and S. Solomon, Power, Levy, Exponential and Gaussian Regimes in
Autocatalytic Financial Systems, cond-mat/0008026, to appear in Eur. Phys.
J. B and \newline
"Finite market size as a source of extreme wealth inequality and market
instability" to appear in Physica A. \newline
Y. Ijiri and H. A. Simon, Skew Distributions and the Sizes of Business Firms
(North-Holland, Amsterdam, 1977). \newline
H. Kesten, Acta Math. 131 (1973) 207. \newline
M. Levy, S. Solomon Power Laws are Logarithmic Boltzmann Laws \newline
International Journal of Modern Physics C , Vol. 7, No. 4 (1996) 595;
adap-org/9607001 \newline
M. Levy M, and S. Solomon (1997), Physica A 242, 90. \newline
M. Levy, H. Levy and S. Solomon, "Microscopic Simulation of Financial
Markets; from Investor Behavior to Market Phenomena" Academic Press, New
York, 2000. \newline
M. Levy, N. Persky, and S. Solomon, "The Complex Dynamics of a Simple Stock
Market Model" International Journal of High Speed Computing, 8, 1996 \newline
A.J. Lotka, (editor) Elements of Physical Biology, Williams and Wilkins,
Baltimore, 1925; \newline
O. Malcai, O. Biham and S. Solomon, Phys. Rev. E, 60, 1299, (1999). \newline
R. Mantegna and H. E. Stanley, An Introduction to Econophysics: Correlations
and Complexity in Finance Cambridge University Press, Cambridge, 1999. 
\newline
M. Marsili, S. Maslov and Y-C. Zhang, Physica A 253,(1998) 403. \newline
S Maslov, "Price fluctuations from the order book perspective - empirical
facts and a simple model" Physica A, present issue. We thank the author for
discussing with us the results prior to publication. \newline
E. W. Montroll Social Dynamics and the Quantifying of Social Forces" , Proc.
Nat. Acad. Sci. USA, Vol 75 No 10, Oct, 1978. \newline
V. Pareto, Cours d''economie politique. Reprinted as a volume of Oeuvres
Compl`etes (Droz, Geneva, 18961965). V. Pareto, Cours d'Economique Politique
(Macmillan, Paris, 1897), Vol. 2. V. Pareto, Le Cours d' ' Economie
Politique (Macmillan, London, 1897). \newline
S. Redner, Am. J. Phys. 58, 267 (1990) ; Eur. Phys. J. B4, 131 (1998). 
\newline
P. Richmond, Power Law Distributions and Dynamic behaviour of Stock Markets,
to appear in Eur. J. Phys 2001. \newline
P. Richmond and S. Solomon, cond-mat/0010222, submitted to J. Quan. Finance. 
\newline
Nadav M. Shnerb, Yoram Louzoun, Eldad Bettelheim, and Sorin Solomon, "The
importance of being discrete: Life always wins on the surface", Proc. Natl.
Acad. Sci. USA, Vol. 97, Issue 19, 10322-10324, September 12, 2000,
http://xxx.lanl.gov/abs/adap-org/9912005l \newline
N. M. Shnerb, E. Bettelheim, Y. Louzoun, O. Agam, S. Solomon, "Adaptation of
Autocatalytic Fluctuations to Diffusive Noise", Phys Rev E Vol 63, No 2,
2001; http://xxx.lanl.gov/abs/cond-mat/0007097 \newline
S. Solomon and M. Levy, adap-org/9609002 , Int. J. Mod. Phys. C7, (1996) 745 
\newline
S. Solomon, in Decision Technologies for Computational Finance, edited by
A.-P. Refenes, A. N. Burgess, and J. E. Moody (Kluwer Academic Publishers,
1998). \newline
S. Solomon, Generalized Lotka-Volterra (GLV) Models and Generic Emergence of
Scaling Laws in Stock Markets, in "Applications of Simulation to Social
Sciences" ,Eds: G Ballot and G. Weisbuch; Hermes Science Publications 2000. 
\newline
S. Solomon and M. Levy, Market Ecology, Pareto Wealth Distribution and
Leptokurtic Returns in Microscopic Simulation of the LLS Stock Market Model; 
{http://arXiv.org/abs/cond-mat/0005416} ; To appear in the Proceedings of
"Complex behavior in economics: Aix en Provence (Marseille), France, 2000". 
\newline
S. Solomon and P. Richmond Stability of Pareto-Zipf Law in Non-Stationary
Economies, http://xxx.lanl.gov/abs/cond-mat/0012479 to appear in the
Proceedings of WEHIA 2000, Marseille. \newline
D. Sornette and R. Cont ,in J. Phys. I France 7 (1997) 431 \newline
H.E. Stanley, L.A.N. Amaral, J.S. Andrade, S.V. Buldyrev, S. Havlin, H.A.
Makse, C.K. Peng, B. Suki and G. Viswanathan, Scale-Invariant Correlations
in the Biological and Social Sciences. Phil. Mag. B, vol. 77, 1998, p. 1373. 
\newline
V. Volterra [1926], Nature, 118, 558. \newline

\end{document}